\documentclass[12pt]{iopart}
\usepackage{graphicx}
\usepackage{iopams}
\usepackage{epstopdf}
\usepackage{hyperref}
\usepackage{cite}
\usepackage{color}

\graphicspath{{./img/}}
\begin{document}
\title{Finite-temperature dynamics of a bosonic Josephson junction}
\author{Y.~M.~Bidasyuk$^{1,*}$, M.~Weyrauch$^1$, M.~Momme$^1$ and O.~O.~Prikhodko$^2$}
\address{$^1$
Physikalisch-Technische Bundesanstalt, Bundesallee 100, D-38116 Braunschweig, Germany}
\address{$^2$
	Department of Physics, Taras Shevchenko National University of Kyiv, Volodymyrska Str. 64/13, Kyiv 01601, Ukraine}
\ead{$^*$Yuriy.Bidasyuk@ptb.de}
\begin{abstract}

In the framework of the stochastic projected Gross-Pitaevskii equation we investigate finite-temperature dynamics of a bosonic Josephson junction (BJJ) formed by a Bose-Einstein condensate of atoms in a two-well trapping potential.
We extract the characteristic properties of the BJJ from the stationary finite-temperature solutions and compare the dynamics of the system with the resistively shunted Josephson model.
Analyzing the decay dynamics of the relative population imbalance we estimate the effective normal conductance of the junction induced by thermal atoms. 
The calculated normal conductance at various temperatures is then compared with predictions of the noise-less model and the model of ballistic transport of thermal atoms.

\end{abstract}

\vspace{2pc}

\submitto{\jpb}

\section{Introduction}

A system of coupled atomic Bose-Einstein condensates (BECs) offers a unique possibility to study at macroscopic scale various quantum coherence phenomena, in particular the bosonic analogue of the Josephson effect \cite{barone}. Bosonic Josephson junctions (BJJ) were realized and investigated in various BEC setups, such as double-well traps \cite{PhysRevA.57.R28,PhysRevLett.84.4521,PhysRevLett.95.010402,levy2007ac}, ring traps or atomic SQUIDs \cite{PhysRevLett.111.205301,PhysRevLett.113.045305}, two-component spinor condensates \cite{PhysRevLett.105.204101,Abad2013}. The rich phase-space portrait of zero-temperature BJJs is explained using a rather simple classical non-rigid pendulum model \cite{PhysRevA.59.620}.
Some studies also go beyond pure Josephson dynamics of the two-well system proposing effective models for treatment of dissipation and fluctuations. These include e.g. finite-temperature damping \cite{PhysRevA.57.R28,lebedev2017exciton,PhysRevA.60.487}, coupling with elementary excitations \cite{PhysRevB.82.195312,PhysRevA.94.033603,doi:10.1002/andp.201700124} or higher modes of the trapping potential \cite{PhysRevA.89.023614}, development of decoherence \cite{PhysRevLett.87.180402,0953-4075-40-10-R01,1367-2630-8-9-189}.

Nevertheless modeling non-equilibrium finite-temperature dynamics of a two-well system in a consistent way poses difficult theoretical problems. 
When two condensates are characterized by different values of the chemical potential then chemical potentials of thermal clouds in each well may also differ from the condensates as well as between them. 
Dissipative dynamics of such a system is governed by two complementary processes characterized in general by different time scales. One process is the incoherent tunneling of thermal atoms through the barrier which leads to equilibration of two thermal clouds located in two wells. The other process is the equilibration of the BEC with the thermal cloud in each well.
A commonly used assumption for such systems  is that 
the second process is much faster than the first one and therefore inside each well the thermal cloud is considered in equilibrium with the condensate\cite{PhysRevA.57.R28,PhysRevLett.113.170601}. The physical picture in this case is analogous to the resistively shunted superconducting Josephson junction, where the current through the junction is represented as a sum of superconducting (Josephson) and normal (Ohmic) current components \cite{PhysRevA.60.487}.

In the present work we analyze the dynamics of the double-well BEC
with the thermal cloud which is internally in equilibrium and the dissipative Josephson dynamics is driven by the equilibration of each of the two BECs with the thermal cloud. This situation is physically relevant for relatively high temperatures, when the average energy of thermal atoms $k_\mathrm{B}T$ is higher than the barrier between two condensates. In this case there is no net normal current as the chemical potential of the thermal cloud is the same on both sides of the barrier. The equilibration of each condensate with the thermal cloud is then similar to the BEC growth process \cite{PhysRevA.62.063609}.
Nevertheless, as we will show, the dynamics of the relative population imbalance between the two condensates again follows the resistively shunted Josephson model with some \textit{effective} value of normal conductance related to the growth rate of the condensate.

We model the finite-temperature dynamics of the partially condensed Bose gas in the framework of the stochastic projected Gross-Pitaevskii equation (SPGPE)  \cite{0953-4075-38-23-008,PhysRevA.81.023630,PhysRevA.77.033616}.
While SPGPE was primarily developed for harmonically trapped condensates, 
it was also successfully applied to trapping potentials with rather large anharmonic part \cite{PhysRevA.88.063620}. However, there are no implementations known to the authors for double-well setups. 
Therefore, testing the SPGPE consistency and applicability for such systems provides a useful extension to the range of problems addressed with this approach. 
Concerning the two equilibration processes mentioned above, the SPGPE approach can not describe the tunneling of thermal atoms through the barrier as it considers the thermal atoms as a static thermal bath. 
On the other hand it can reliably model the equilibration between the BEC and the thermal cloud at relatively high temperatures, which makes it a valid tool to analyze the process under study.

The paper is organized as follows. In the second section we introduce the resistively shunted Josephson model and discuss how the normal current affects the dynamics of otherwise stable Josephson solutions. In the third section we describe the formalism of stochastic Gross-Pitaevskii equation and estimate the parameters required for dynamical simulations. In the section~4 we describe the relation between the two models and analyze the static properties of the bosonic Josephson junction. Finally, in the section~5 we simulate the dynamical behavior of the system.
By analyzing the decay dynamics of the population imbalance between two wells we  extract the values of the effective conductance corresponding to the normal Ohmic current in the resistively shunted Josephson model. This normal conductance can be complemented with the estimates based on the ballistic transport of normal atoms at low temperatures \cite{PhysRevA.57.R28,levy2007ac} providing a general picture of the two-well dissipative dynamics in a wide range of temperatures.

\section{Resistively shunted Josephson model}

Let us consider two trapped BECs weakly coupled by a barrier potential forming a Bosonic Josephson junction (BJJ). The low-energy collective dynamics of such a system is conveniently described using two dynamical quantities: relative population imbalance $Z = (N_1-N_2)/N$ and relative phase $\theta = \theta_1-\theta_2$, where $N_{1,2}$ and $\theta_{1,2}$ are the atom numbers and the phases of each BEC cloud, $N=N_1+N_2$ is the total number of atoms in two condensates. Accounting for weak dissipative effects in the system these quantities obey the following set of equations similar to to the resistively shunted Josephson (RSJ) model \cite{PhysRevA.60.487}:
\begin{eqnarray}
\label{eq:rsj}
\frac{dZ}{dt} &= \omega_\mathrm{J} \sqrt{1-Z^2} \sin\theta - G\frac{\Delta\mu}{\hbar} \\
\frac{d\theta}{dt} &= -\omega_\mathrm{C} Z - \frac{\omega_\mathrm{J} Z}{\sqrt{1-Z^2}} \cos\theta
\label{eq:rsj2}
\end{eqnarray}
where $\omega_\mathrm{J}$ and $\omega_\mathrm{C}$ are the parameters related to the Josephson coupling energy $E_\mathrm{J} = \hbar\omega_\mathrm{J} N/2$ and the capacitive energy $E_\mathrm{C} = 2 \hbar\omega_\mathrm{C}/N $ of the junction. 
The chemical potential difference between two condensates is related to the population imbalance as $\Delta\mu = \hbar \omega_\mathrm{C} Z$.
The second term in (\ref{eq:rsj}) is the analogue of the normal Ohmic current in the Josephson junction with the dimensionless parameter $G$ as the normal conductance (or the conductance of the shunt resistor connected in parallel with the Josephson junction).

If $\omega_\mathrm{J} \ll \omega_\mathrm{C}$ and $E_\mathrm{C} \ll E_\mathrm{J}$ then the system is considered to be in the Josephson regime \cite{RevModPhys.73.307}. In this case and without dissipation ($G=0$) equations~(\ref{eq:rsj}) and (\ref{eq:rsj2}) support two types of solutions. 
First, oscillations of $Z$ and $\theta$ around zero mean, which are known as Josephson plasma oscillations. Second, small amplitude oscillations of $Z$ around a non-zero mean with uniformly growing $\theta$, which are often called running-phase solutions or macroscopic quantum self-trapping (MQST) states.

\begin{figure}[htbp]
\centering
\includegraphics[width=0.6\linewidth]{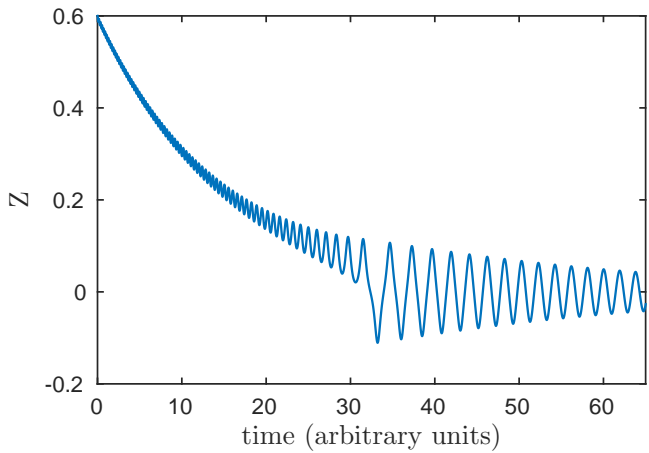}
\caption{Example solution of the RSJ equations. First part of the evolution time shows the decay of MQST state and the second part shows the decay of plasma oscillations.
}\label{fig:je_z}
\end{figure}

When dissipation is included ($G>0$) both plasma oscillations and MQST states decay exponentially with time. An example of such dynamical behavior is presented in the figure~\ref{fig:je_z}. The system is initially in a decaying MQST state
 (decaying mean value of the oscillations) 
but when the population imbalance $Z$ reaches values close to zero it switches to the plasma oscillations around zero mean but with decaying amplitude.
With the assumptions that the system is in the Josephson regime ($\omega_\mathrm{J} \ll \omega_\mathrm{C}$) and the normal conductance is small ($G \ll 1$) the decay rates of plasma oscillations and the mean value of $Z$ in MQST states can be derived from (\ref{eq:rsj}) and (\ref{eq:rsj2}) as
\begin{equation}
\tau_\mathrm{MQST}^{-1} = G\omega_\mathrm{C}, \qquad \tau_\mathrm{Plasma}^{-1} = \frac12 G\omega_\mathrm{C}.
\label{eq:decay_rates}
\end{equation}

The normal conductance $G$ is commonly associated with finite-temperature dissipative effects in the system. However, if we intend to analyze the temperature dependence of this quantity from the measurements of the decay rates, then it is necessary to account for possible temperature dependence of $\omega_\mathrm{C}$ as well. This will be addressed in the section~4.

\section{Stochastic Projected Gross-Pitaevskii equation}

In order to analyze dissipative processes in a two-well system without any free parameters we model the problem using the approach of Stochastic Projected Gross-Pitaevskii equation (SPGPE). 
As a model system we consider here the experimental setup of Ref.~\cite{levy2007ac}.
The system is characterized by the mean field Gross-Pitaevskii Hamiltonian operator $H_\mathrm{GP}$:
\begin{equation}
H_\mathrm{GP}\, \psi(\mathbf{r},t) = \left[-\frac{\hbar^2\nabla^2}{2m} + V_\mathrm{ext}(\mathbf{r},t) + g |\psi(\mathbf{r},t)|^2 \right] \psi(\mathbf{r},t).
\label{eq:gpham}
\end{equation}
with the nonlinear interaction parameter $g=4\pi\hbar^2 a/m$, where $a$ is the $s$-wave scattering length of $^{87}\mathrm{Rb}$ and $m$ is the atom mass.
The potential $V_\mathrm{ext}$ consists of the static cylindrically symmetric harmonic trap and a movable gaussian barrier splitting the system into two wells along the long axis of the trap \cite{levy2007ac}:
\[
V_\mathrm{ext}(\mathbf{r},t) = V_0(\mathbf{r}) + V_\mathrm{b}(\mathbf{r},t)
\]
where
\[
V_0(\mathbf{r}) = \frac{m}{2}\left[\omega_r^2(x^2+y^2) + \omega_z^2 z^2\right], \qquad V_\mathrm{b}(\mathbf{r},t) = U_\mathrm{b} e^{-2[x-x_\mathrm{b}(t)]^2/w_\mathrm{b}^2}.
\]
The parameters of the trap and the barrier potentials are chosen in correspondence with the experimental setup  \cite{levy2007ac}: $\omega_r/2\pi=224~\mathrm{Hz}$, $\omega_z/2\pi=26~\mathrm{Hz}$, $w_\mathrm{b}=0.7~\mathrm{\mu m}$, $U_\mathrm{b}/h = 3~\mathrm{kHz}$ (the barrier height was not measured in the experiment and we choose here the value that provide comparable estimate of the Josephson critical current). 
The barrier position $x_\mathrm{b}(t)$ defines the driving protocol for creation of the initial population imbalance. 
In the experiment \cite{levy2007ac} the barrier position was fixed and the harmonic trap center was moved from the initial shift of $0.7~\mathrm{\mu m}$ (from the barrier center) to zero within a certain time $\tau$. 
For the SPGPE calculations it is necessary that the harmonic trapping potential is time-independent, therefore we model the same process by shifting the barrier position by the same distance within the same time. We choose the time of the barrier shift $\tau=5\,\mathrm{ms}$ to be well inside the AC Josephson regime according to the results of \cite{levy2007ac}, which means that it produces a pronounced chemical potential difference between two wells and drives the system into a MQST state.

Classical field or $C$-field methods are based on the concept of splitting the many-particle system into highly occupied low-energy modes described by the coherent classical field $\psi(\mathbf{r},t)$ and sparsely occupied incoherent high-energy modes forming a thermal bath. 
Such splitting is 
conveniently represented in the basis of single-particle eigenstates $\phi_n$ of the harmonic trapping potential $V_0$
\[
\left[-\frac{\hbar^2\nabla^2}{2m} + V_0(\mathbf{r})\right]\phi_n = e_n\phi_n.
\]
The classical field $\psi(\mathbf{r},t)$ is then a coherent superposition of these states with energies below the chosen cut-off energy $e_\mathrm{cut}$
\begin{equation}
\psi (\mathbf{r},t) = \sum\limits_{n\in C} c_n(t) \phi_n(\mathbf{r}), 
\qquad C = \{n:e_n \leq e_\mathrm{cut}\}
\label{eq:oscrep}
\end{equation}
Such a classical field obeys the Stochastic projected Gross-Pitaevskii equation  \cite{0953-4075-38-23-008,PhysRevA.81.023630,PhysRevA.77.033616} (in \cite{PhysRevA.77.033616} it is referred to as a ``simple growth'' SPGPE as it neglects some additional scattering terms reported there):
\begin{equation}
d\psi (\mathbf{r},t) = -\frac{i}{\hbar} \mathcal{P} H_\mathrm{GP} \psi(\mathbf{r},t) dt + \frac{\gamma}{\hbar} \mathcal{P} (\mu - H_\mathrm{GP}) \psi(\mathbf{r},t) dt + dW(\mathbf{r},t) 
\label{eq:spgpe}
\end{equation}
where $\mathcal{P}$ is a projection operator to the $C$-space.
\[
\mathcal{P} \psi (\mathbf{r},t) = \sum\limits_{n\in C} \phi_n(\mathbf{r}) \int d\mathbf{r}' \phi_n^*(\mathbf{r}') \psi (\mathbf{r}',t).
\]
The noise term $dW(\mathbf{r},t)$ in (\ref{eq:spgpe}) is the Gaussian complex noise with the correlation
\begin{equation}
\langle dW^*(\mathbf{r}',t) dW(\mathbf{r},t) \rangle = \frac{2 \gamma}{\hbar \beta}  \delta_C(\mathbf{r}',\mathbf{r}) dt,
\label{eq:noise}
\end{equation}
with $\beta=1/k_BT$ and $\delta_C(\mathbf{r}',\mathbf{r}) = \mathcal{P}\delta(\mathbf{r} - \mathbf{r}') =  \sum_{n\in C} \phi_n(\mathbf{r}) \phi_n^*(\mathbf{r}')$ being the projection of the $\delta$ function to the $C$-space. The stochastic dynamics described by (\ref{eq:spgpe}) and (\ref{eq:noise}) is analogous to a complex-valued Wiener process.

The coefficient $\gamma$ in (\ref{eq:spgpe}) defines the growth rate of the condensate. It can be derived from the kinetic theory with an assumption that the above-cutoff particles behave as an ideal Bose gas \cite{PhysRevA.77.033616}. The resulting expression yields:
\begin{equation}
\gamma = \gamma_0 \sum_{j=1}^{\infty} \frac{e^{\beta\mu(j+1)}}{e^{2\beta e_\mathrm{cut}j}} \Phi [e^{\beta(\mu-e_\mathrm{cut})},1,j]^2.
\label{eq:gamma}
\end{equation}
with $\gamma_0 = 4m a^2 k_\mathrm{B} T/\pi\hbar^2$ and $\Phi$ is the Lerch transcendent.

\subsection{Stationary states and estimation of SPGPE parameters}

A crucial part of SPGPE calculation is the estimation of parameters.
This means that by specifying the total number of particles in the system $N_\mathrm{T}$ (which in our case is chosen according to \cite{levy2007ac} as $N_\mathrm{T} = 1\times10^5$) and the temperature $T$ we need to define the chemical potential of the system $\mu$ and the cut-off energy $e_\mathrm{cut}$.
In the existing implementations \cite{PhysRevA.81.023630,PhysRevA.62.063609} this is done by constructing and analyzing stationary distributions of the condensate and thermal atoms within the Hartree-Fock (HF) approximation. Let us briefly outline this procedure as it was implemented for the present study.

In the static HF approximation \cite{griffin2009bose,0953-4075-41-20-203002} the condensate wave function is defined as a solution of the stationary GPE
\begin{equation}
\mu \psi_0(\mathbf{r}) = 
\left( -\frac{\hbar ^2 }{2m} \nabla^2
+ V_{\mathrm{ext}}(\mathbf{r}) + g \left[|\psi_{0}(\mathbf{r})|^2  
+ 2\tilde n_{0}(\mathbf{r}) \right] \right)\psi_0 \;, 
~\label{eq:zng_sgpe}
\end{equation}
and the stationary thermal particles density 
\begin{equation}
\tilde n_0(\mathbf{r}) = \int \frac{d\mathbf{p}}{(2\pi\hbar)^3} f_\mathrm{BE}(\mathbf{r},\mathbf{p})
\label{eq:zng_eq_n}
\end{equation}
is obtained by integrating the Bose-Einstein distribution function
\begin{equation}
f_\mathrm{BE}(\mathbf{r},\mathbf{p}) = \frac{1}{e^{\beta(E(\mathbf{r},\mathbf{p})-\mu)}-1},
\label{eq:bedist}
\end{equation}
where the energy of thermal atoms in the effective HF potential is defined as
\[
E(\mathbf{r},\mathbf{p}) = \frac{\mathbf{p}^2}{2m} +  V_\mathrm{ext}(\mathbf{r}) + 2g[|\psi_{0}(\mathbf{r})|^2 + \tilde n_0 (\mathbf{r})],
\]
which in turn contains dependence on both $\psi_{0}(\mathbf{r})$ and $\tilde n_0 (\mathbf{r})$. The above expression for energy is also used to define the HF density of states for the system
\begin{equation}
\rho(\epsilon) = \int \frac{d\mathbf{r} d\mathbf{p}}{(2\pi\hbar)^3} \delta(\epsilon - E(\mathbf{r},\mathbf{p})).
\label{eq:dens_states}
\end{equation}

The two coupled equations (\ref{eq:zng_sgpe}) and (\ref{eq:zng_eq_n}) can be numerically  solved self-consistently with the additional constraint of the fixed total number of atoms in the system
\[
N_\mathrm{T} = N + \tilde N = \int d\mathbf{r} |\psi_{0}(\mathbf{r})|^2 + \int d\mathbf{r} \tilde n_0 (\mathbf{r}).
\]
As a result we get the equilibrium value of the chemical potential $\mu$, the stationary particle distributions of the condensate and thermal atoms, and the  number of atoms in the condensate $N$ and in the thermal cloud $\tilde N$.

The cut-off energy $e_\mathrm{cut}$ is obtained from the condition that the $C$-field includes only highly occupied modes. In practice this means that from the Bose distribution we find the maximal energy with specified occupation $n_\mathrm{cut}$ (in present work we choose $n_\mathrm{cut}=1$)
\[
e_\mathrm{cutHF} = k_\mathrm{B} T \ln\left(1+\frac1{n_\mathrm{cut}}\right) + \mu.
\]
Since for the SPGPE we use an oscillator basis, we require that the number of states below $e_\mathrm{cutHF}$ is equal to the number of oscillator states below $e_\mathrm{cut}$ \cite{PhysRevA.81.023630}:
\[
\int_0^{e_\mathrm{cutHF}} d\epsilon \rho(\epsilon) = \int_{e_0}^{e_\mathrm{cut}} d\epsilon \rho_\mathrm{HO}(\epsilon),
\]
where $e_0 = \hbar(2\omega_r+\omega_z)/2$, $\rho_\mathrm{HO}(\epsilon) = \epsilon^2/(2\hbar^3\omega_r^2\omega_z)$ are the harmonic oscillator ground state energy and the density of states, respectively, defined for our cylindrically symmetric trap. 

\begin{figure}[htbp]
\centering
\includegraphics[width=0.75\linewidth]{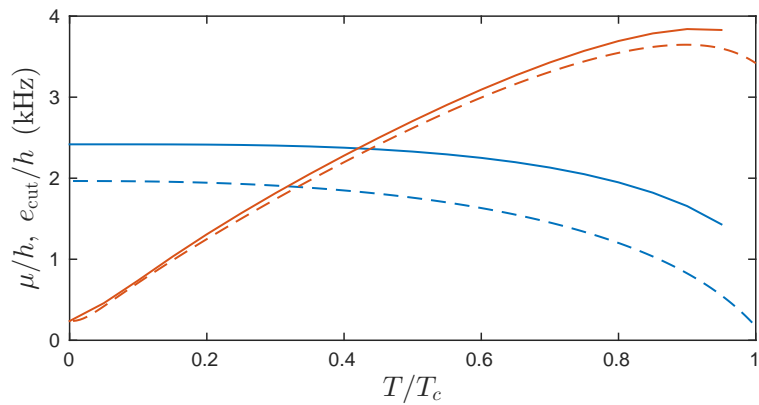}
\caption{Equilibrium chemical potential (blue lines) and the cut-off energy (red lines) as a function of temperature as obtained from the full static Hartree-Fock approximation (solid lines) and with the Tomas-Fermi approximation (dashed lines). Temperature is shown relative to the condensation temperature $T_c=225~\mathrm{nK}$.
}\label{fig:ecut}
\end{figure}

The procedure described above can be further simplified using the Thomas-Fermi approximation. Then the equations (\ref{eq:zng_sgpe}) and (\ref{eq:zng_eq_n}) are solved semi-analytically. Additionally, only the harmonic part of the trapping potential is considered and this allows also to get analytical expressions for the density of states. Such an approach is used in most of the existing implementations of SPGPE \cite{0953-4075-38-23-008,PhysRevA.81.023630,PhysRevA.77.033616,PhysRevA.62.063609}. In order to check the applicability of this approximation in our system we compare in figure~\ref{fig:ecut} the values of $e_\mathrm{cut}$  and $\mu$ obtained with the full numerical solution of (\ref{eq:zng_sgpe}) and (\ref{eq:zng_eq_n}) and with the Thomas-Fermi approximation. One may see that accurate treatment of the effective potential with the static HF approach significantly changes the equilibrium chemical potential estimate. The main reason for this discrepancy is that analytical expressions obtained in the Thomas-Fermi approximation do not account for the barrier and only consider the harmonic part of the trap. Quite surprisingly, for the cut-off energy both estimates yield similar results.

The obtained dependencies $e_\mathrm{cut}(T)$ and $\mu(T)$ allow us to define the range of temperatures where SPGPE can be used. One of the SPGPE applicability criteria is that the above-cutoff states are well approximated by the unperturbed oscillator states, which requires $\mu \ll e_\mathrm{cut}$. Also the barrier potential should be reliably represented in the basis of single particle states, which imposes additional requirements $U_b \ll e_\mathrm{cut}$ and $w_\mathrm{b} \ll R_\mathrm{cut}=\sqrt{2e_\mathrm{cut}/m\omega_r^2}$ \cite{PhysRevA.88.063620}. This means that reliable results can be expected only for rather limited range of temperatures $T\gtrsim0.6T_c$. In the present work we choose therefore to limit our calculations to the temperature range $0.6T_c \leq T \leq 0.9 T_c$. On the boundaries of this region we get $e_\mathrm{cut}(0.6T_c) = 14.18\,\hbar\omega_r$ and $e_\mathrm{cut}(0.9T_c) = 18.56\,\hbar\omega_r$. The resulting number of modes in $C$-region is $4546$ and $9904$ respectively.

\begin{figure}[htbp]
\centering
\includegraphics[width=0.6\linewidth]{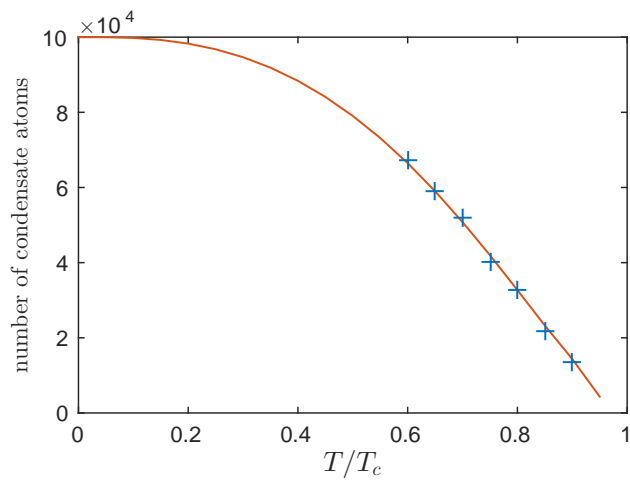}
\caption{The number of condensate atoms from the static HF approximation (solid line) and from SPGPE simulations (crosses)
}\label{fig:nc}
\end{figure}

In order to check the consistency of the SPGPE model with the defined parameters we verify if the equilibrium number of condensate atoms in the SPGPE simulation matches with the value from the static HF solutions. This is done by evolving the equation (\ref{eq:spgpe}) in time with a static barrier potential. Then the condensate fraction can be calculated from the Penrose-Onsager criterion \cite{PhysRev.104.576} stating that 
the largest eigenvalue of the one-body density matrix provides an estimate of the condensate atom number.
The one-body density matrix is constructed using  the ergodicity hypothesis to replace ensemble average by the time average of a single trajectory \cite{PhysRevA.72.063608,PhysRevA.77.033616}:
\[
\rho_C(\mathbf{r},\mathbf{r}') = \langle \psi^*(\mathbf{r},t) \psi(\mathbf{r}',t) \rangle_t
\]
where $\langle ... \rangle_t$ denotes the time average. In practice such averaging was done from 100 snapshots of the $C$-field taken uniformly from a single SPGPE run over the time interval of $0.05~\mathrm{s}$, which corresponds to roughly $12$ transverse trap periods.
The atom number obtained with this approach can then be directly compared with the prediction of the static HF approximation (see figure~\ref{fig:nc}) justifying the choice of SPGPE parameters as well as the overall consistency of the algorithm.

\section{Static properties of BJJ at finite temperatures}

Before we investigate the dynamical behavior of the system let us first show how the static HF approximation (\ref{eq:zng_sgpe},\ref{eq:zng_eq_n}) can be used to estimate the parameters of the Josephson model and to analyze the static properties of the BJJ.
A standard zero-temperature Gross-Pitaevskii equation (GPE) for the two-well system can be reduced to the system of Josephson equations (\ref{eq:rsj},\ref{eq:rsj2}) (without the dissipative term) using a two-mode approximation \cite{PhysRevLett.84.4521}. To this end the condensate wave function is  represented as a coherent superposition of two solutions $\psi_1$ and $\psi_2$ localized (mainly) in each well with time-dependent amplitudes and phases:
\begin{equation}
\psi(\mathbf{r},t) = \sqrt{N_1(t)}e^{i\theta_1(t)}\psi_1(\mathbf{r}) + \sqrt{N_2(t)}e^{i\theta_2(t)}\psi_2(\mathbf{r}).
\label{eq:tm_ansatz}
\end{equation}
This leads to the system of two Josephson equations (\ref{eq:rsj}) and (\ref{eq:rsj2}) without normal current term ($G=0$). In the case of symmetric double-well system and assuming weak coupling between two wells the other parameters of the Josephson equations acquire simple expressions in terms of localized solutions $\psi_{1,2}(\mathbf{r})$ (assuming them being orthogonal and normalized to unity): 
\begin{equation}
\omega_\mathrm{J} = -\frac{2}{\hbar}\int d\mathbf{r} \psi_1 \left[-\frac{\hbar^2\nabla^2}{2m} + V_\mathrm{ext} + g N|\psi_2|^2 \right] \psi_2
\end{equation}
 and 
\begin{equation}
\omega_\mathrm{C} = \frac{gN}{\hbar}\int d\mathbf{r}|\psi_1|^4 = \frac{gN}{\hbar}\int d\mathbf{r}|\psi_2|^4,
\label{eq:omc1}
\end{equation}
 with $N$ as the total number of condensate atoms.

In order to see how the normal conductance appears from the dissipative finite-temperature equation we consider for simplicity the SPGPE (\ref{eq:spgpe}) without the noise term. This ``silent'' version of SPGPE is formally equivalent to the Gross-Pitaevskii equation with phenomenological damping \cite{PhysRevA.57.4057}:
\begin{equation}
i \hbar \frac{\partial\psi}{\partial t} = (1-i\gamma) (H_\mathrm{GP}-\mu) \psi.
\label{eq:dgpe}
\end{equation}
This equation with the two-mode ansatz (\ref{eq:tm_ansatz}) reduces to the Josephson equations (\ref{eq:rsj},\ref{eq:rsj2}) including the normal current term with the conductance defined as $G=\gamma$.
However one has to keep in mind that in equation (\ref{eq:dgpe}) the damping parameter $\gamma$ is considered to be a purely phenomenological parameter.

In practical applications the value of the interaction parameter $\omega_\mathrm{C}$ obtained from (\ref{eq:omc1}) leads to a poor agreement with the results of dynamical GPE calculations \cite{1367-2630-13-3-033012,PhysRevA.87.053625}. Therefore an improved two-mode model has been proposed \cite{PhysRevA.87.053625} that effectively accounts for the variations of the shape of the localized solutions by introducing a linear approximation to the integral
\[
\frac{\int d\mathbf{r} |\psi_0|^2 |\psi_{\Delta N}|^2}{\int d\mathbf{r} |\psi_0|^4} = 1 - \alpha \frac{\Delta N}{N}
\]
where $\psi_0$ and $\psi_{\Delta N}$ are unity-normalized condensate ground state solutions corresponding to  $N$ and $N+\Delta N$ atoms respectively. This results in a simple rescaling of the parameter $\omega_\mathrm{C}$:

\begin{equation}
\omega_\mathrm{C} = (1-\alpha) \frac{gN}{\hbar}\int d\mathbf{r}|\psi_{1,2}|^4.
\label{eq:omc2}
\end{equation}
More details on this model can be found in \cite{PhysRevA.87.053625,Nigro2017}. We adapt this procedure for the finite-temperature case by replacing the GPE solution by the static HF condensate solution thus repeatedly solving the system of equations (\ref{eq:zng_sgpe},\ref{eq:zng_eq_n}) with small variations in the number of atoms at each temperature. 
We find a value for the coefficient $\alpha\approx0.3$ that is almost independent on the temperature and closely corresponds to the value derived analytically in \cite{PhysRevA.87.053625} for a three-dimensional condensate. Alternatively, the same result can be obtained by numerically evaluating the capacitive energy as $E_\mathrm{C} = 4 \partial\mu/\partial N$ \cite{RevModPhys.73.307,1367-2630-10-4-045009}, however we find that the first approach gives more numerically stable results in the finite-temperature case.

\begin{figure}[htbp]
\centering
\includegraphics[width=0.5\linewidth]{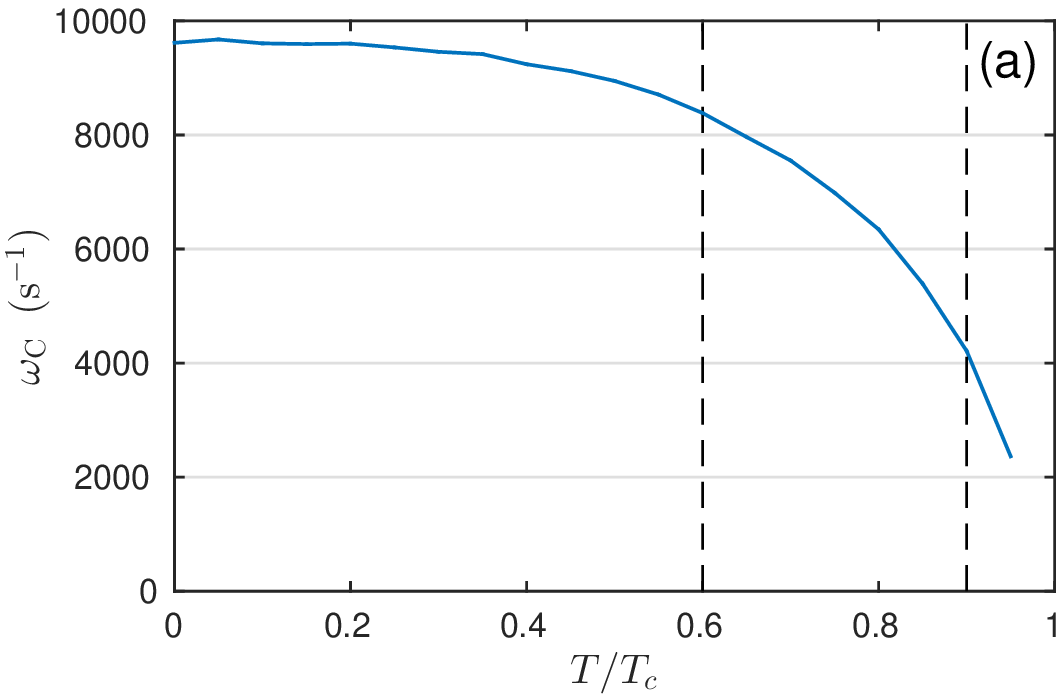}\includegraphics[width=0.5\linewidth]{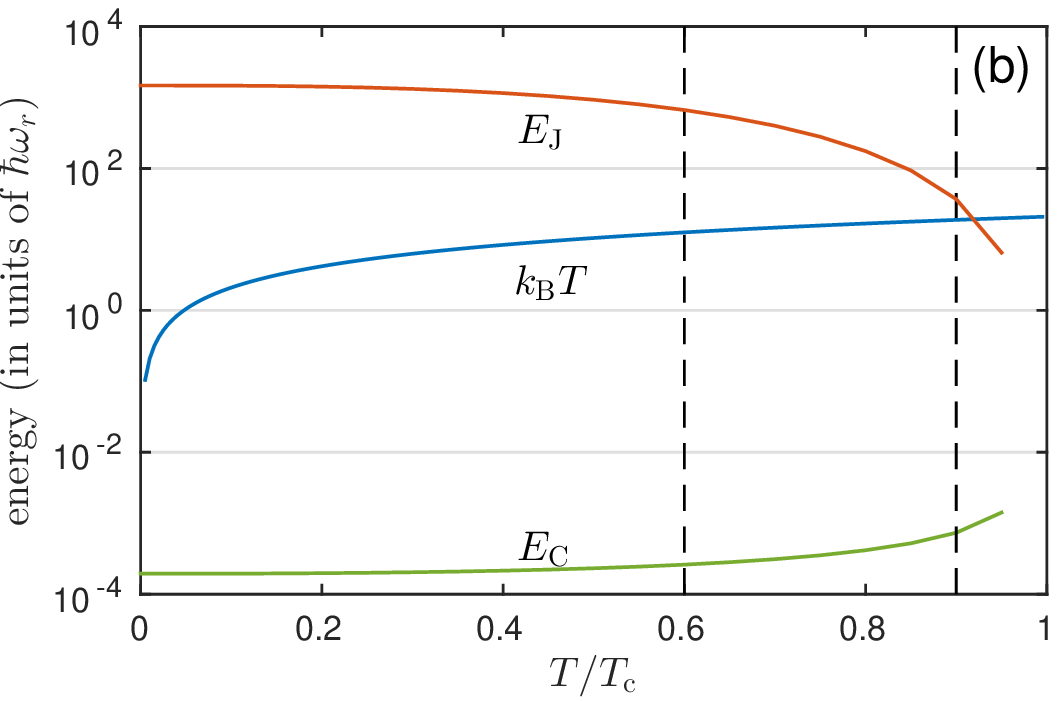}
\caption{Temperature dependence of various characteristic quantities of BJJ calculated from the static HF solutions. Panel (a) shows the temperature dependence of $\omega_\mathrm{C}$. Panel (b) shows the temperature dependence of the Josephson coupling energy $E_\mathrm{J}$ (red line) and the capacitive energy $E_\mathrm{C}$ (green line) and compares them to $k_\mathrm{B}T$ (blue line). Note the logarithmic scale of the vertical axis of panel (b). Vertical dashed lines on both panels mark the region of temperatures used for SPGPE simulations.
}\label{fig:omc}
\end{figure}

The resulting dependence $\omega_\mathrm{C}(T)$ is presented in figure~\ref{fig:omc}(a). We see that in the low-temperature region the obtained numbers agree well with the value $\omega_\mathrm{C} \approx 9000~\mathrm{s}^{-1}$ reported in \cite{levy2007ac}. With growing temperature $\omega_\mathrm{C}$ decays rather rapidly, mainly due to the depletion of the condensate particle number. 
Within the temperature region considered in the present study this parameter changes from approximately $8000~\mathrm{s}^{-1}$ (for $T=0.6T_\mathrm{c}$) to $4000~\mathrm{s}^{-1}$ (for $T=0.9T_\mathrm{c}$). 
Such strong variation indicates that it is absolutely crucial to take into account temperature variations of this quantity when the two-mode approximation is used for finite-temperature BJJ.

The physically relevant quantities related to the parameters $\omega_\mathrm{J}$ and $\omega_\mathrm{C}$ are the Josephson coupling energy $E_\mathrm{J} = \hbar\omega_\mathrm{J} N_\mathrm{c}/2$ and the capacitive energy $E_\mathrm{C} = 2 \hbar\omega_\mathrm{C}/N$. In the zero-temperature case the requirement $E_\mathrm{C} \ll E_\mathrm{J}$ ensures that the coherence of two condensates is not destroyed by quantum phase fluctuations. In a finite temperature case thermal fluctuations are of more importance and the coherence requirement reads $k_\mathrm{B} T \ll E_\mathrm{J}$ \cite{PhysRevA.57.R28,PhysRevLett.87.180402,0953-4075-40-10-R01,1367-2630-8-9-189}. In figure~\ref{fig:omc}(b) we compare these three characteristic energies of the system. 
We see that at high temperatures close to $0.92 T_\mathrm{c}$ two energies become close $k_\mathrm{B} T \approx E_\mathrm{J}$, which means that the two-well system becomes partially incoherent. This loss of coherence affects the observability of coherent Josephson oscillations between two wells. However, in the self-trapped state any net coherent current is suppressed and equilibration of the two condensates takes place only due to incoherent processes. Therefore, such partial loss of coherence should not affect the decay rate of MQST state and the normal conductance which we intend to determine.

\section{Dissipative dynamics of MQST states}

\begin{figure}[htbp]
\centering
\includegraphics[width=\linewidth]{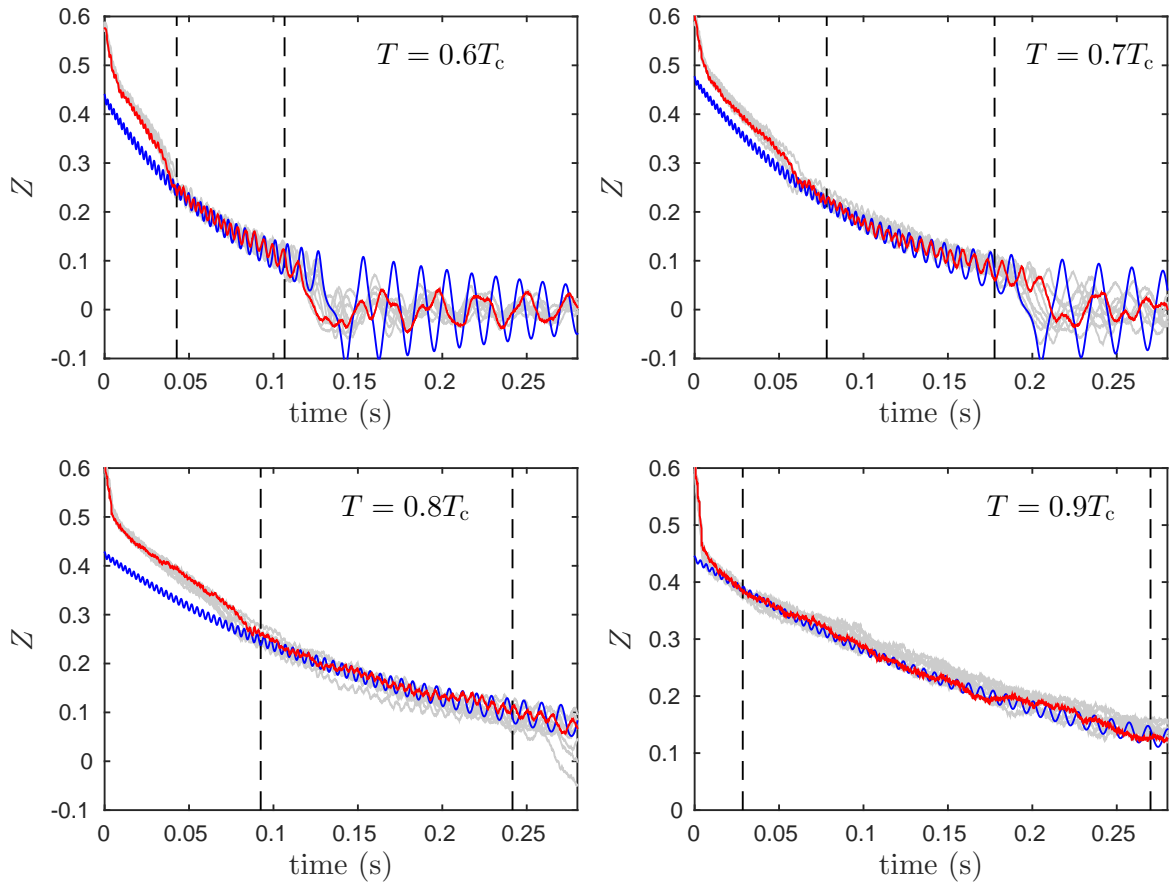}
\caption{Dynamics of the population imbalance $Z$ corresponding to the decaying MQST state at four different temperatures. Red line on each panel is the result of one SPGPE run, blue lines --- solutions of the Josephson equations. Light gray lines on the background show the results of all other SPGPE runs with the same temperature. Dashed vertical lines mark the regions used to extract the decay rates.
}\label{fig:zt}
\end{figure}

The dynamics of the system is modeled by numerically evolving Eq.~(\ref{eq:spgpe}) in time.
The simulation is started in a trap with a barrier position shifted by $0.7~\mathrm{\mu m}$ off the trap center. The system is first evolved in time with a barrier kept static until it reaches a thermalized state. (which takes approximately $0.5~\mathrm{s}$ of the evolution time). Thermalization is detected by the saturation of the condensate particle number. 
The barrier is then shifted within $5~\mathrm{\mu s}$ to the center of the harmonic trap. 
The evolution of the system is simulated for another $0.3~\mathrm{s}$ after the trap symmetry is restored.
For each temperature we make 10 independent SPGPE runs.
From each SPGPE run we extract the dependence $Z(t)$ by integrating on every time step the coordinate-space solution over regions spanned by each potential well. 
Figure~\ref{fig:zt} shows these dependencies obtained at different temperatures (time $t=0$ corresponds to the beginning of the barrier movement).

Let us first analyze the obtained $Z(t)$ time series, and in particular their difference to the solutions of the Josephson equations.
The initial sharp drop of the population imbalance within first $5\,\mathrm{\mu s}$ is due to the barrier move. Afterwards we observe decaying MQST state similar to the figure \ref{fig:je_z}. This part of the evolution is qualitatively well reproduced by the RSJ model (\ref{eq:rsj},
\ref{eq:rsj2}).
The amplitude of the Josphson oscillations is however noticeably lower than that obtained from RSJ model calculations. This is mainly due to the fact that populations obtained in SPGPE simulations effectively contain all modes of the $C$-region, not only the condensate mode. The other feature observed in the simulations, making them different from the prediction of RSJ model is an additional ``kink'' of the decay rate at $Z\approx 0.3$. This additional dissipative effect is persistent across multiple SPGPE runs and for all temperatures up to $0.8T_\mathrm{c}$. This effect is likely a result of a resonant generation of one of the low-energy collective modes of the trap. A similar dissipative effect was studied in \cite{PhysRevA.94.033603} however for a quite different toroidal geometry of the trapping potential.

\begin{figure}[htbp]
\centering
\includegraphics[width=0.85\linewidth]{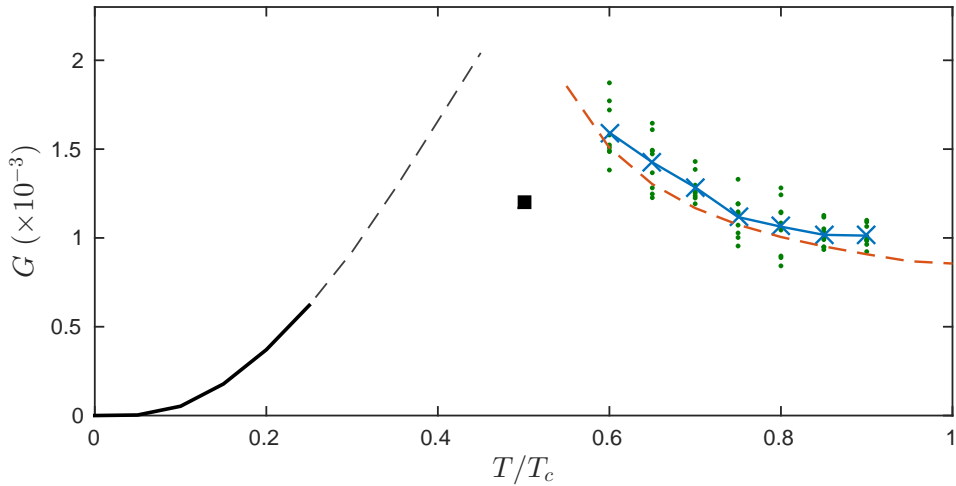}
\caption{Temperature dependence of the normal conductance in the bosonic Josephson junction. Shown are the results of SPGPE simulations (green dots show the results extracted from individual SPGPE runs, blue crosses show averaged values), the damping parameter $\gamma(T)$ from the equation (\ref{eq:gamma}) (dashed red line), and the estimate of Eq.~(\ref{eq:ak}) (solid black line in the region where Arrhenius-Kramers formula is applicable, dashed black line is the extrapolation to higher temperatures). Black square shows the conductance value extracted from the experiment \cite{levy2007ac}.
}\label{fig:decay_rate}
\end{figure}

In order to calculate the decay rate $\tau_\mathrm{MQST}^{-1}$ we make an exponential fit to the part of $Z(t)$ dependence that correspond to the decaying MQST state.
For the exponential fit we choose the region of $Z(t)$ after the ``kink'' mentioned above and before the transition to the plasma oscillations. 
Using the values of $\omega_\mathrm{C}(T)$ calculated from the static HF approximation we can now extract the effective normal conductance from (\ref{eq:decay_rates}) as $G = \tau_\mathrm{MQST}^{-1}/\omega_\mathrm{C}$. The results are presented in figure~\ref{fig:decay_rate}. These values can be compared to the values of damping parameter $\gamma$ defined by (\ref{eq:gamma}), which represent the normal conductance in the ``silent'' version of the model. We see that the ``noisy'' dynamics introduces on average only a small bias to the conductance, which may be as well due to some small inaccuracy in the $\omega_\mathrm{C}$ calculation. One may also see that the spread of the values obtained from the individual SPGPE runs is rather small and averaging over only 10 runs provides a reliable approximation.
Therefore we can conclude that the decay rate of the MQST Josephson state is 
only weakly influenced by thermal noise and can be reasonably reproduced by noise-less dissipative model. This result is similar to the decay of dark soliton states in one-dimensional condensates analyzed in \cite{PhysRevLett.104.174101}, where  the lifetime of such states was shown to be also almost insensitive to the thermal noise. However, for other states, e.g. quantum vortices, noisy and noise-less dynamics provide the decay rates that are different by an order of magnitude \cite{PhysRevA.81.023630}.

It is worth noticing, that at the highest considered temperature $T=0.9T_\mathrm{c}$ the Josephson oscillations are not observable any more in the $Z(t)$ dependence (see lower right panel of figure~\ref{fig:zt}) due to a partial loss of coherence in the system. Nevertheless, the extracted normal conductance still shows a value close to the noise-less approach.

The proposed approach to estimate the normal conductance in the system is limited to relatively high temperatures.
It is also instructive to compare the obtained results with the model of the ballistic transport of thermal atoms \cite{PhysRevA.57.R28,levy2007ac}. This model provides a basic order-of-magnitude estimate of the normal conductance at low temperatures $k_\mathrm{B}T < U_\mathrm{b}-\mu$. 
It uses the Arrhenius-Kramers formula to estimate the crossing rate of a thermal particle through the barrier:
\[
P_n = \frac{\omega_r}{2\pi} e^{-\frac{U_\mathrm{b}-\mu}{k_\mathrm{B}T}}.
\]
Then the total conductance with our notation can be expressed as \cite{PhysRevA.57.R28}:
\begin{equation}
G = \frac{\hbar P_n \tilde N}{k_\mathrm{B} T N_T} =\frac{1}{k_\mathrm{B}T}\frac{\hbar \omega_r}{2\pi} \frac{\tilde N}{N_T} e^{-\frac{U_\mathrm{b}-\mu}{k_\mathrm{B}T}}.
\label{eq:ak}
\end{equation}
This formula can be easily evaluated using the values of $\tilde N$ and $\mu$ from the static HF approximation (see figure~\ref{fig:decay_rate}).

Let us discuss the results of the two considered models of normal conductance in the system.
The ballistic conductance defined by Eq.~(\ref{eq:ak}) is due to the tunneling of thermal atoms through the barrier. 
It estimates the rate at which two thermal clouds with different chemical potentials come to an equilibrium.
Therefore the estimate of the MQST decay rate by Eq.~(\ref{eq:ak}) relies on the assumption that the condensate and the thermal cloud in each well remain in mutual equilibrium. 
The SPGPE model on the other hand models the equilibration between the thermal cloud and two condensates. It does not reflect any transport of thermal atoms through the barrier as the chemical potential of the thermal bath remains uniform in the system. Extrapolating the predictions of both models to intermediate temperatures we see that they show similar values around $T\approx 0.5 T_c$. If different equilibration processes take place simultaneously, then the combined decay rate will be lower then provided by any of them separately. Therefore we can consider these extrapolations to provide an upper bound on the effective normal conductance in the system. This is clearly seen if the obtained model values are compared to the result from \cite{levy2007ac} where the MQST decay rate was measured at the temperature $T=0.5 T_c$, which is outside the validity regions of both ballistic estimate and SPGPE. As is seen in figure~\ref{fig:decay_rate} extrapolation of both models overestimates the value of normal conductance at this point giving only an order-of-magnitude estimate. 

Another peculiar property of the normal conductance which follows from our estimates is that its temperature dependence appears to be non-monotonic. Indeed, the low-temperature estimate gives a monotonically growing temperature dependence while the SPGPE results show a decline with temperature. This suggests the existence of a maximum of the normal conductance in the intermediate temperature region. Such non-monotonic behavior may be a consequence of an interplay between two complementary processes: equilibration of the thermal clouds in two wells and equilibration of the thermal clouds with the condensates. Alternatively, other phenomena such as noise-enhancement of stability \cite{PhysRevE.72.061110} may be considered to describe such behavior. More detailed theoretical and experimental verification of this phenomenon will be a subject for a future work.

\section{Conclusions}

We have implemented the formalism of the stochastic projected Gross-Pitaevskii equation for a system of two weakly coupled atomic BECs.  
We show how the relevant characteristics of a finite-temperature bosonic Josephson junction can be extracted from the static Hartree-Fock approximation and used to analyze the dynamical SPGPE results.

Using the developed implementation we have analyzed the decay dynamics of a self-trapped state in a bosonic Josephson junction.
Our calculations conform with the RSJ model where the thermal effects are encapsulated in the normal Ohmic contribution to the total particle current. The corresponding normal conductance appears to be noise insensitive and completely described by the damping coefficient $\gamma$ defined within the SPGPE model. 
For high temperatures close to  $T_\mathrm{c}$ the development of thermal decoherence is observed as a reduction or complete suppression of the Josephson oscillations between the two condensates. Nevertheless, the effective normal  current still agrees well with the prediction of the noise-less model.

The results are compared with the qualitative estimate of ballistic transport model. These comparison suggests that two processes of thermalization (between two thermal clouds and between thermal cloud and BEC) may have similar characteristic time scales which leads to the non-monotonic temperature dependence of the normal conductance. Such an effect is expected to be traceable in the existing experimental setups.

\section*{Acknowledgements}

The authors are thankful to N. P. Proukakis for fruitful discussions and comments on the manuscript.

\section*{References}
\bibliographystyle{unsrt}
\bibliography{refs}
\end{document}